\documentclass[twocolumn,showpacs,preprintnumbers,amsmath,amssymb]{revtex4}

\usepackage{graphicx}
\usepackage{bm}

\begin{document}

\title{When do generalized entropies apply? How phase space volume determines entropy}

\author{Rudolf Hanel$^{1}$ and  Stefan Thurner$^{1,2}$}

\affiliation{
$^1$ Section for Science of Complex Systems, Medical University of Vienna, Spitalgasse 23, A-1090, Austria\\
$^2$  Santa Fe Institute, 1399 Hyde Park Road, Santa Fe, NM 87501, USA
} 

\begin{abstract}
We show how the dependence of phase space volume $\Omega(N)$ of a classical system on its size $N$ 
uniquely determines its extensive entropy. We give a concise criterion when this entropy is not of 
Boltzmann-Gibbs type but has to assume a {\em generalized} (non-additive) form.  We show that generalized 
entropies can only exist when the dynamically (statistically) relevant fraction of degrees of freedom in the system 
vanishes in the thermodynamic limit. These are systems where the bulk of the degrees of freedom is frozen 
and is practically statistically inactive. Systems governed by generalized entropies are therefore systems whose 
phase space volume effectively collapses to a lower-dimensional 'surface'. We explicitly illustrate the 
situation for binomial processes and argue that generalized entropies could be relevant for self organized 
critical systems such as sand piles,  for spin systems which form meta-structures such as vortices, domains, 
instantons, etc., and for problems associated with anomalous diffusion. 
\end{abstract}

\pacs{05.20.-y, 02.50.Cw, 05.90.+m}

\maketitle


Entropy relates the number of states of a system to an {\em extensive} quantity, which plays a fundamental role in its  thermodynamical description. 
Extensive means that when two initially isolated systems $A$ and $B$ -- with $\Omega_A$ and $\Omega_B$ the respective numbers of states -- are brought in contact, 
the entropy of the combined system $A+B$ is $S(\Omega_{A+B}) = S(\Omega_A) + S(\Omega_B)$. 
Extensivity  is not to be confused with {\em additivity} which is the property that $S(\Omega_A \Omega_B) = S(\Omega_A) + S(\Omega_B)$. 
Both, extensivity and additivity coincide if  the number of states in the combined system is $\Omega_{A+B}=\Omega_A\Omega_B$.
Clearly,  for a non-interacting system Boltzmann-Gibbs (BG) entropy, $S_{\rm BG}[p]=\sum_i^\Omega  g_{\rm BG}(p_i)$,  
with $g_{\rm BG}(x)=- x\ln x$, is simultaneously extensive {\em and} additive.
By 'non-interacting'  systems (short-range, ergodic,  mixing, Markovian, ...) we mean  $\Omega_{A+B}=\Omega_A\Omega_B$. 
For interacting statistical systems this is in general not true. If  phase space is only partly visited this means $\Omega_{A+B} < \Omega_A\Omega_B$. In this case,   
it may happen that an additive entropy (such as BG) no longer is extensive and vice versa. 
With the hope to understand  {\em interacting} statistical systems within a thermodynamical formalism 
and to ensure extensivity of entropy, so called {\em generalized entropies}  have been introduced which usually assume trace form
\begin{equation}
 	S_{\rm gen}[p]=\sum_{i=1}^\Omega g(p_i)\quad, \qquad [\Omega \,\,... \,\,{\rm number\,\,  of \,\, states}]
\label{S_g} 
\end{equation} 
where $g$ is some function of $p$. 
 It has been shown that $g$ can not assume any functional form, but generalized entropies of trace form are 
restricted  to  the  family of functions 
 \begin{equation}
	 S_{c,d}[p] \propto \sum_{i=1}^\Omega \Gamma(d+1,1-c\log p_i) \quad , 
 \label{gamma}
 \end{equation}
$\Gamma(.,.)$ being the incomplete gamma function, 
whenever a minimum set of requirements on $g$ hold \cite{Hanel2011}.  These requirements are
the first three of the four Shannon-Khinchin (SK) axioms \cite{Shannon1948,Kinchin1957},
SK1:  Entropy $S$ depends continuously on $p$ ($g$ is continuous),  
SK2: entropy is maximal for the equi-distribution $p_i=1/\Omega$ ($g$ is concave.
In physical systems this represents the equi-partition principle in micro-canonical ensembles),
SK3: adding a zero-probability state to a system, $\Omega+1$ with $p_{\Omega+1}=0$,  does not change the entropy ($g(0)=0$), and 
SK4: the entropy of a system -- composed of sub-systems $A$ and $B$ -- equals the entropy of $A$ plus the expectation value of the entropy 
of $B$, conditional on $A$. 
If SK1-SK4 hold, the only possible entropy is BG \cite{Shannon1948,Kinchin1957}.  If only SK1-SK3 hold (additivity axiom violated) Eq. (\ref{gamma}) is the 
generalized entropy with the constants $(c,d)$ characterizing the universality class of entropy. $(c,d)=(1,1)$ is the class of BG entropy, 
 $(c,d)=(q,0)$ is the class of Tsallis entropies. 
 A universality class $(c,d)$ not only characterizes the entropy of the system completely in the thermodynamic limit, it also specifies its distribution functions. 
 Many recently introduced generalized entropic forms appear to be  special cases of Eq. (\ref{gamma}) \cite{Hanel2011}. The associated distribution 
 functions are
\begin{equation}
	{\cal E}_{c,d,r}(x)=  e^{    - \frac{d}{1-c}  \left[ W_k \left( B \left(1-\frac{x}{r} \right)^{ \frac{1}{d} } \right)  -  W_k(B)  \right]  }\, ,
\label{gexp}
\end{equation}
with  $B\equiv \frac{(1-c)r}{1-(1-c)r} \exp \left(  \frac{(1-c)r}{1-(1-c)r} \right) $, and as one possible choice,
 $r=(1-c +cd)^{-1}$, \cite{Hanel2011}. 
The function $W_k$ is the $k$'th branch of the Lambert-W function, which is a solution of the equation
$x=W(x)\exp(W(x))$. Only branch $k=0$ and branch $k=-1$ have real solutions $W_k$.
Branch $k=0$ is necessary  for all classes with $d\geq 0$, branch $k=-1$ for $d<0$.  The generalized logarithm for 
the entropy Eq. (\ref{gamma}) is the inverse of ${\cal E}=\Lambda^{-1}$. 
Further properties of systems where SK1-SK3 hold are reported in \cite{htmgm11}.
  
It has often been argued that for statistical systems with strong and  long-range correlations, Boltzmann-Gibbs statistical mechanics  
loses its applicability, and that under these circumstances generalized entropies become necessary. This is certainly not true in 
general. While correlations can be the reason for non-Boltzmann distribution functions, BG entropy often 
remains the correct extensive entropy of the system \cite{Hanel2009}.

In this paper we clarify the conditions under which BG entropy breaks down as the extensive entropy of a system. 
For ergodic systems, covering phase space,  BG is always valid, regardless of what the correlations in the system might be. 
This was explicitly shown for binary systems in \cite{Hanel2009}. 
It is obvious that the {\em structure} of phase space, i.e. Gibbs $\Gamma$-space,  is responsible  
for the Boltzmann-Gibbs framework to collapse and for generalized entropies to become necessary. 
Here we show that mere non-ergodicity is not enough:  for  generalized entropies to become necessary, $\Gamma$-space has to 
collapse in a specific way. 

In the following we derive all results in terms of growth of phase space volume as a function of system size. 
We illustrate our results for binary systems where a graphical representation is possible in terms of decision trees. Binary systems 
with correlations \cite{definetti, jaynes}
have been studied in the light of generalized entropies in  
\cite{TsallisGellMannSato2005, Marsh2006,Moyano2006,RodriguezSchwammleTsallis2008,Hanel2009}. 
On the basis of growth of $\Gamma$-space as a function of the number of states we present a set of concise 
criteria when generalized entropies are unavoidable and specify them by their universality classes. 


What does extensivity mean?
Consider a system with $N$ elements, each of which can be in one of $m$ states. The number of system configurations 
(microstates) are denoted by $\Omega(N)$, which depends on $N$ in a system-specific way. 
Starting with Eq. (\ref{S_g})  for equi-distribution, $p_i=1/\Omega$ (for all $i$), we have  $S_g=\sum_{i=1}^{\Omega} g(p_i) = \Omega g(1/\Omega)$.  
Extensivity for two subsystems $A$ and $B$ means that
\begin{equation}
	\Omega_{A+B}g\left( 1/\Omega_{A+B}  \right) = \Omega_{A}g\left( 1/ \Omega_{A}  \right) + \Omega_{B}g\left( 1/ \Omega_{B}  \right) \quad .
\label{dum1}
\end{equation}
Using the primary scaling property of generalized entropies 
$\lim_{x\to0^+} \frac{g(\lambda x)}{g(x)} = \lambda ^c$, (see \cite{Hanel2011}), 
we get asymptotically $g'(x) = c g(x)/x$, and Eq. (\ref{dum1}) becomes
\begin{equation}
	g'\left( 1/ \Omega_{A+B}  \right) = g' \left( 1/ \Omega_{A}  \right) + g'\left( 1/ \Omega_{B}  \right) \quad .
\end{equation}
The derivative of $g$ is the  generalized logarithm, 
$g'(x)=-\Lambda(x)$, and 
\begin{eqnarray}
	\frac{1}{\Omega_{A+B}} = {\cal E} \left[  \Lambda  \left( \frac{1}{\Omega_{A}}  \right) 
	+ \Lambda \left( \frac{1}{\Omega_{B}}  \right)  \right]  = \frac{1}{\Omega_{A}} \otimes_g \frac{1}{\Omega_{B}}
	\quad . 
\end{eqnarray}
A generalized product $\otimes_g$ can now be  defined as
$x \otimes_g y \equiv   {\cal E} \left[  \Lambda (x)  + \Lambda(y) \right]$.
If each 'particle' can be in one of $m$ states, we finally get for the number of states in the system  
\begin{equation}
	\frac{1}{\Omega(N)} = {\cal E} \left[  N \Lambda \left( \frac1m \right) \right] \quad, 
\label{om}
\end{equation}
or if we use the distribution functions and generalized logarithms of generalized entropies, Eq. (\ref{gexp}), 
the number of microstates grows asymptotically as  
\begin{equation}
	\Omega(N)   =    \frac{1}{  {\cal E}_{c,d}(\mu (c-1) N)} =   \exp \left[  \frac{d}{1-c} W_k \left( \mu (1-c) N^{\frac1d} \right) \right] \quad , 
\label{grow}
\end{equation}
where  $\mu$ is some positive constant.  
At this stage note that for all non-BG systems, i.e. $(c,d)\neq (1,1)$, 
the number of states $\Omega(N)$ grows  sub-exponentially with $N$.

Inversely, given the phase space volume as a function of system size, we can now  compute the generalized entropy, i.e. its 
universality class characterized by $(c,d)$. 
Again using the primary scaling property for generalized entropies and de L'Hospital rule
$\lambda ^c = \lim_{x\to0^+} \frac{g(\lambda x)}{g(x)} =  \frac{\lambda g'(\lambda x)}{g'(x)} = \frac{\lambda \Lambda(\lambda x)}{\Lambda(x)} $,
together with Eq. (\ref{om}) we get for large $\Omega(N)$
\begin{equation}
	\lambda^{c-1} \Lambda  \left( 1/ \Omega(N)  \right) = \Lambda \left(  \lambda / \Omega(N)  \right) \quad , 
\end{equation}
or $\lambda^{\frac{1}{c-1}}  = \lim_{N\to \infty} \frac{ \Omega(\lambda N)}{\Omega(N)} $, which can be simplified to 
\begin{equation}
	\frac{1}{1-c}= \lim_{N \to \infty} N \frac{ \Omega'(N)}{\Omega(N)} \quad .
	 \label{cformel}
\end{equation}
For $d$ we use the secondary scaling relation for generalized entropies \cite{Hanel2011}, 
$(1+a)^d=\lim_{x \to 0} \frac{g(x^{1+a})}{x^{ac}g(x)}$. Taking the derivative with respect to $a$ on both sides 
we get 
\begin{equation}
	d(1+a)^{d-1} = (1+a)^d  \lim_{x\to 0} \log x  \left( \frac{x^{1+a}g'(x^{1+a})}{ g(x^{1+a})}  - c \right) .
\end{equation}
Set $a\to 0$, and use de L'Hospital rule to get $d= \lim_{x\to 0}   (1-c + x g''(x)/g'(x) )$ so that   with 
 Eq. (\ref{om}) we have 
\begin{equation}
	d=  \lim_{N \to \infty}  \log  \Omega  \left( \frac1N \frac{\Omega}{ \Omega'} +c-1 \right) \quad .
\label{dformel}
\end{equation}


\begin{figure}
\includegraphics[width=6.0cm]{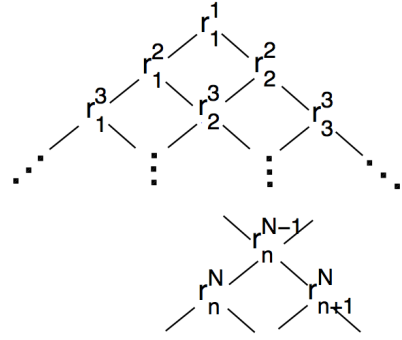}
\caption{Decision tree (triangle) for binary processes representing the probabilities $r^N_n$  of $n$ heads 
and $N-n$ tails after $N$ throws. $r^N_n$ is the probability of a  {\em specific}   sequence, 
$\{\varphi_1,\varphi_2,\dots,\varphi_N\}$, which does not depend on the order of events, but only on the 
number of $n$ events in state $\varphi_i=1$ and $N-n$ events in state $\varphi_j=0$. Leibnitz rule (scale invariance) 
\cite{TsallisGellMannSato2005} holds if  $r^N_n + r^N_{n+1}= r^{N-1}_n$  at all levels $N$ and $N-1$. }
\label{leibnitz}
\end{figure}

To see how  phase space  collapses for generalized entropy systems 
we illustrate the above result in the context of binary sequences, 
e.g. $\varphi=\{0,1,1,0,0,1,\cdots\}$.
Let $\left.\varphi\right|_N=\{\varphi_i\}_{i=1}^N$ denote a sequence of length $N$. 
Any correlations in sequences (on all levels of $N$)  \cite{Hanel2009} are completely determined by the 
set of joint probability functions
$\{ p_N\}_{N=1}^\infty$, where $p_N(\varphi_1,\varphi_2,\dots,\varphi_N)$ is the 
joint probability of a  sequence of length $N$.
A sequence $\left.\varphi\right|_N$ contains $k(N)=\sum_{i=1}^N \varphi_i$ 'ones'
and $N-k(N)$ 'zeros'. 
If  $p_N(\varphi_1,\varphi_2,\dots,\varphi_N)$ is totally symmetric in its arguments the 
probability of sequences of length $N$ depends on $k$ only, $r^N_k=p_N(\varphi_1,\varphi_2,\dots,\varphi_N)$.
There exist ${N}\choose{k}$ sequences $\left.\varphi\right|_N$ with $k$ 'ones' and $N-k$ 'zeros' and 
$\sum_{k=0}^N {N \choose{k}} r^N_k=1$. 
The $r^N_k$ can be arranged into a triangle, Fig. \ref{leibnitz},
representing the equi-probable sequences of binary events in a 'decision tree'. 
The full phase space volume (on level $N$) is 
\begin{equation}
	\Gamma_N=\{0,1 \}^N   \qquad , \qquad \Omega(N) =  |\Gamma_N| = 2^N   \quad .
\end{equation}
In other words, the number of  
microstates (number of sequences up to level $N$) is $\Omega(N)=2^N$. 
Obviously the number of states increases exponentially and when substituted in Eqs. (\ref{cformel}) and (\ref{dformel})
we recover $(c,d)=(1,1)$, i.e. BG entropy. 

We now introduce restrictions to phase space such that not all sequences are allowed on all levels $N$ anymore. 
We denote the number of states in the restricted phase space by  $\Omega^{(R)}(N) = |\Gamma_N^{(R)}| $, and can immediately discuss an  
interesting  fact. Imagine that phase space of sequences is extremely confined, say to a situation where in the thermodynamic limit 
all sequences approach a common point (same number of 'ones' in the sequence) 
 $\xi=\lim_{N\to \infty} k(N)/N$. In all cases where this point $\xi$ is neither $0$ or $1$, 
BG is the only possible extensive entropy. This can be formulated in a\\

{\bf Theorem:} Define a restricted phase space $\Gamma^{(R)}_N$ on level $N$ by 
	\[
	\Gamma^{(R)}_N\equiv\left\{ 
		\left.\varphi\right|_N\in\Gamma_N \;\;:\;\; \underline{K}(N) \leq \sum_{i=1}^N \varphi_i \leq \overline{K}(N)
		\right\}\quad,
\]
i.e., at level $N$ the system only allows sequences $\left.\varphi\right|_N$ with more than 
$\underline{K}(N)$ and less than $\overline{K}(N)$ 'ones'. 
If the restriction of phase space is such that $\lim_{N\to \infty}  \frac{\underline{K}(N)}{N} =  \lim_{N\to \infty}  \frac{\overline{K}(N)}{N} =\xi$, where $\xi \in (0,1)$, 
then asymptotically the number of states grows exponentially ($\Omega(N)= b^N$ for some  $b>0$),  
and the extensive entropy is BG. \\

The proof is to show that both, lower and upper bounds for $\Omega(N)$, yield BG.
The theorem states that for generalized entropies to exist it is necessary 
that the sequences are constrained to the situation where either $\lim_{N\to \infty}  \frac{\overline{K}(N)}{N} =0$, or 
$\lim_{N\to \infty}  \frac{\underline{K}(N)}{N} =1$. In other words the sequences are asymptotically confined to a region 
of measure zero around the flanks of the decision triangle, i.e. the {\em boundary} of phase space. 
The theorem has two further implications: 

1. In case of probability distributions $p_N$ which are {\em not} totally symmetric in their arguments, generalized entropies can
exist even though phase space need not be limited to the boundary of the decision triangle (as in the theorem). 
If the number of sequences $\varphi\in\Gamma^{(R)}$ (i.e. the number of free decisions) up to level $N$ 
grows sufficiently sub-linearly with $N$, then the limit-points of sequences may be found along the entire base of the decision 
triangle (compare remark on super-diffusive random walks below). This means that the multiplicity of sequences
with $k$ out of $N$ 'ones' in the large $N$ limit grows sufficiently slower than the binomial multiplicity 
for totally symmetric  $p_N$.

2. The theorem can trivially be generalized from binary processes to $m$-state systems.  
This is done by  passing from the binomial to a multinomial description.


\begin{figure}
\includegraphics[width=2.8cm]{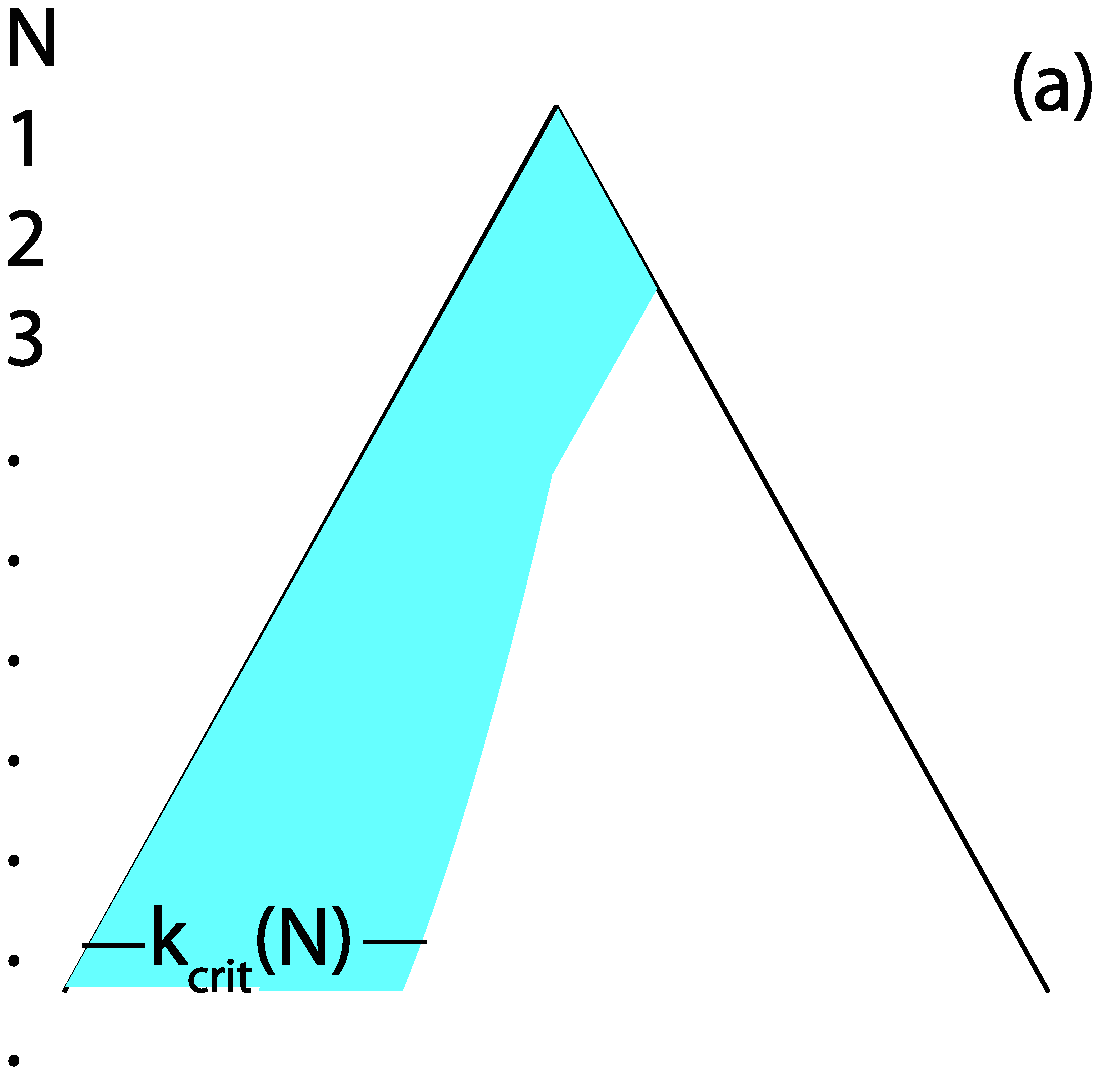}
\includegraphics[width=2.8cm]{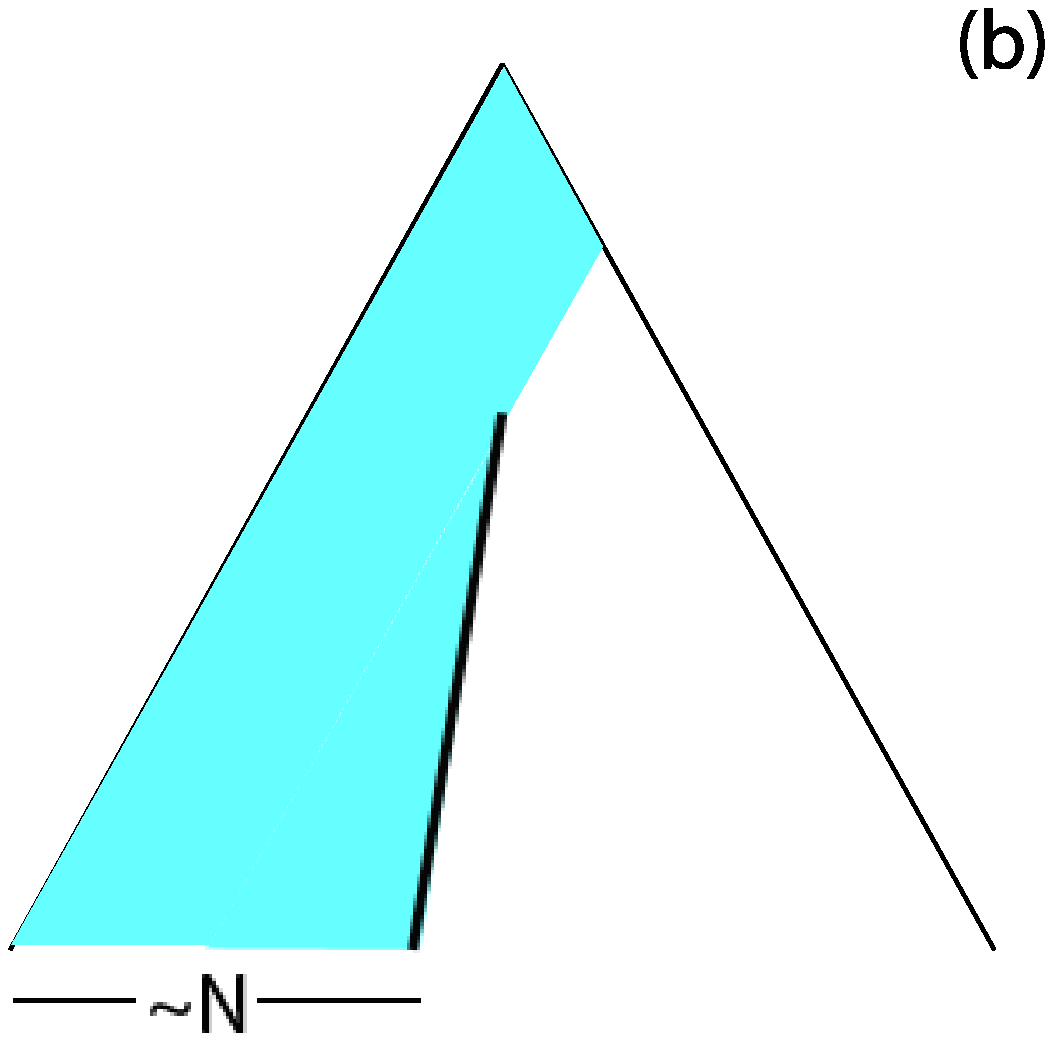}
\includegraphics[width=2.8cm]{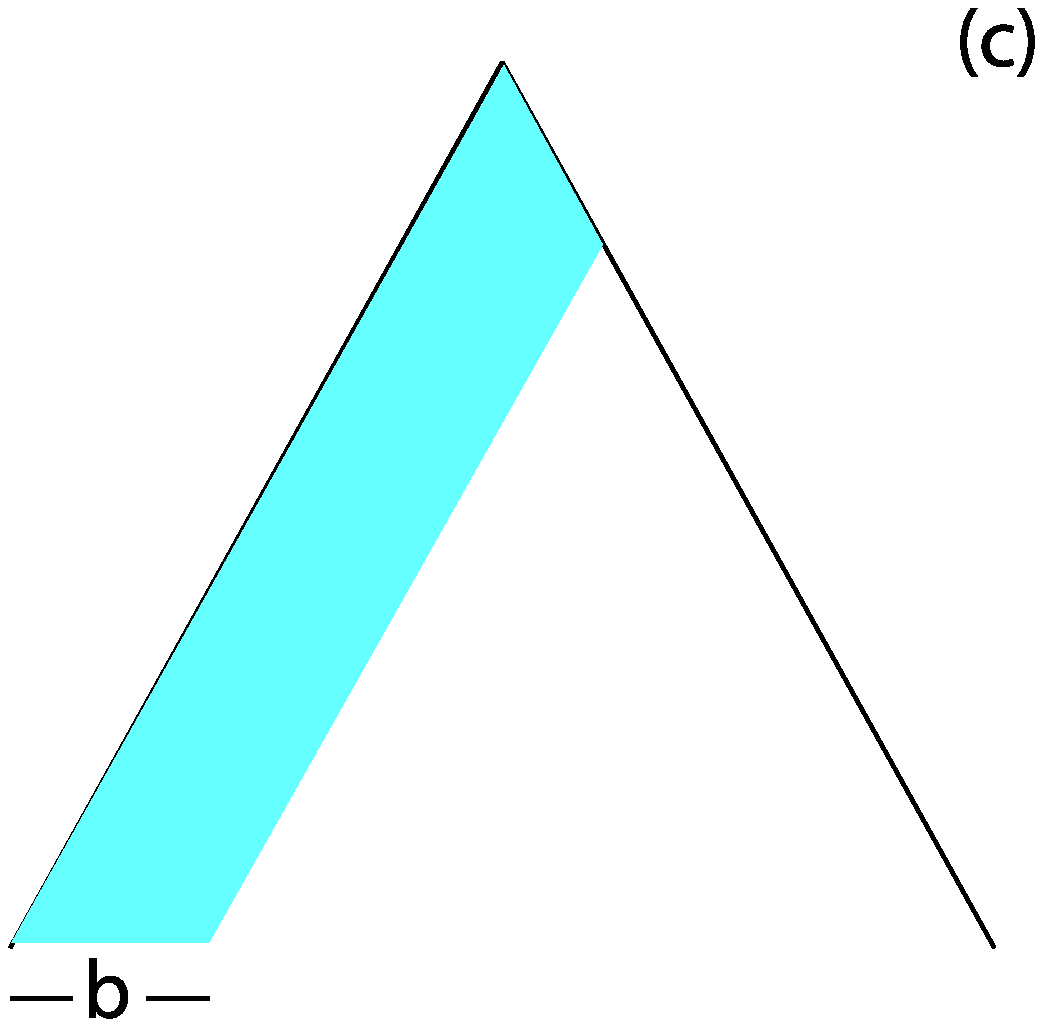}
\caption{Binary decision trees: (a)  Schematic view of  allowed sequence regions. If sequences  are confined to the shaded regions 
(left of critical sequence line $k^{\rm crit}$)  the extensive entropy of the system is not BG but a generalized entropy. 
(b) Maximum line $k^{\rm crit}$ at which BG entropy  starts. Any system containing this line or sequences to the right of it, will be BG. 
(c) If the region is confined to a strip of size $b$, the extensive entropy is Tsallis entropy, $S_{q,0}$, with $q=1-\frac1b$.  }
\label{dreieck}
\end{figure}

We now show how different restrictions on phase space  lead to various  specific generalized entropies. 
We assume the existence of a critical  sequence $\varphi^{\rm crit}$ which follows the path $k^{\rm crit}(N)$, 
see Fig. \ref{dreieck} a. This means that after $N$ steps the sequence has produced a maximum of  $k^{\rm crit}(N)$ 'ones'.
To the right of this sequence all sequences are forbidden. The phase space volume  of such systems grows like 
\footnote{For $m$ states 
	 	the critical number is obtained similarly:
		$\Omega^{(R)}(N) = \sum_{j=1}^{k^{\rm crit} (N) } \sum_{|n|=j} \binom{N}{ n* }$,
		with the multi-indices $n=(n_1,n_2\dots,n_{m-1})$ and $n^*=(n_1,n2, \dots, n_{m-1},N-|n|)$, with $|n|=\sum_{i=1}^{m-1}n_i$.
		The multinomial factor is  $\binom{N}{n^*}=N!/\prod_{j=1}^m n^*_j!$.
	        } 
\begin{equation}
	\Omega^{(R)}(N) = \sum_{i=1}^{k^{\rm crit} (N) }\binom{N}{i} \quad .
\label{crit}
\end{equation}
For any $k$,  
$\lim_{N \to \infty} \sum_{i=1}^{k^{\rm } }  \binom{N}{i} /  \binom{N}{k^{\rm }  } =1$, 
which allows to asymptotically approximate Eq. (\ref{crit}) 
\begin{equation}
	\Omega^{(R)}(N) \approx  \binom{N}{k^{\rm crit} (N) } \quad .
\end{equation}
Using Stirling's formula, taking logs on both sides and keeping terms to  leading order we arrive at  
\begin{equation}
	k^{\rm crit} (N) \approx N \exp \left[  W_{-1} \left( - \frac1N \log \Omega(N)  \right)  \right]    \quad .
\label{res}
\end{equation}
This means that for any system whose sequences are confined to regions left to the critical sequence 
$k^{\rm crit} (N)$, generalized entropies as specified in Eq. (\ref{grow}) are necessary. We now discuss some examples. \\


{\bf Maximum restricted phase space}. Consider $(c,d)=(1,1)$, i.e.  $\Omega(N)=2^N$. From Eq. (\ref{res}) we get 
$k^{\rm crit} (N) \approx N$. This means that for systems with generalized entropies $k^{\rm crit} (N)$ grows in a 
sufficiently sub-linear way with $N$, e.g. $k^{\rm crit}(N) \propto N^\alpha$ with $0<\alpha<1$.  If $\alpha=1$ and 
$k^{\rm crit} (N)= \varepsilon N$, no matter how small $\varepsilon>0$, the system belongs to BG.

{\bf Power-law growth}.
For a power-like growth of phase space, $\Omega(N)=N^b$, we have 
$k^{\rm crit} (N) \approx N \exp [  W_{-1} ( \frac bN \log \frac bN -\frac bN \log b  )  ] $. Expanding the Lambert-W function
we get $k^{\rm crit} (N) \approx b \exp(-\frac{\log b}{1-\log \frac bN}) \to  b$, in the large $N$ limit. The phase space 
collapse is seen in the decision triangle as a restriction to a strip of width $b$, Fig. \ref{dreieck} c. In this case  
$(c,d)=(1-\frac 1b,0)$, i.e. Tsallis entropy applies exactly. This is a well known result  
\cite{TsallisGellMannSato2005,Tsallis_GellMann_Book}.

{\bf Streched exponential growth}.
For stretched exponential growth $\Omega(N)=\exp(\lambda N^{\gamma})$,  Eq. (\ref{res}) can be rewritten to 
$k^{\rm crit} (N) \approx \log \Omega / W_{-1} [-\log \Omega /N]$ and the Lambert-W term is reasonably approximated 
by $\log(N/ \log \Omega) + \log(\log (N/ \log \Omega))$. With this  
$k^{\rm crit} (N) \approx \frac{ \lambda }{1-\gamma} \frac{N^{\gamma}}{\log (N)}$, and the entropy is $(c,d)=(1,1/\gamma)$. \\

Note that systems with confined areas in their decision trees are examples for  strong memory. 
The system has to remember how many  'ones' have occurred in its trajectories, see examples in \cite{Moyano2006,Marsh2006}.
Inversely, given a critical sequence line $k^{\rm crit} (N)$,  the universality class of the 
corresponding generalized  entropy $(c,d)$ can be computed. 


Given the dependence of phase space volume $\Omega$ on system size we  showed how to determine the associated 
extensive generalized entropy  by computing the exponents $(c,d)$. We demonstrated that different generalized (non-additive) 
entropies -- i.e. $(c,d)\neq (1,1)$ -- correspond to different ways of sub-exponential growth of $\Gamma$-space.  We related the 
growth of phase space volume to the increase of the number of {\em statistically relevant} (dynamical) micro-states in the system. 
We found that whenever the fraction of dynamical variables $\frac kN$ vanishes  for large $N$,  $\lim_{N\to \infty} \frac kN=0$, 
generalized entropies become unavoidable.  This extreme confinement of relevant variables  to a set of measure zero means that 
almost all states in the system are the same, or equivalently, the bulk  of the degrees of freedom is frozen. In other words, 
statistically relevant activity happens within a tiny fraction, $\frac kN$, of the system which can be seen as a collapse of 
phase space volume to some low-dimensional 'surface'.

In conclusion we hypothesize that generalized entropies are relevant for physical systems being dominated by 'surface effects', 
including the following:\\
$\bullet$ {\em Self organized critical systems}.  In sandpiles consider discrete sites where  sand grains can be. The (binary) 
state of a site is being occupied by a grain or not. In a sandpile the bulk of the system is occupied and just the surface of the 
pile contains its statistically relevant degrees of freedom. The trajectory of a sand  grain in a classical sandpile model follows 
sequences much alike those in the decision tree, Fig. \ref{dreieck} c.  
\\
$\bullet$ {\em Spin systems with dense meta-structures}, such as spin-domains, vortices, instantons, caging, etc. 
If these meta-structures bind a vast majority of spins into (metastable) objects,  
the remaining spins -- not belonging to these structures -- can move freely only in surface-like regions between these objects. 
For instance spin systems on random networks growing with constant connectedness  
(number of links divided by number of nodes squared) can be shown to require Tsallis entropy.
\\
$\bullet$ {\rm Super-diffusion}. Consider a one dimensional accelerating random walk, where each left-right decision is 
followed by $N^\beta$ ($0<\beta<1$) steps in that direction ($N$ being the total
number of steps the walk has so far taken). This process is a super-diffusive process 
($\langle x^2(t)\rangle \propto t^{2-\beta}$) which requires a generalized entropy of type $(c,d)=(1,1/\beta)$. 
\\
$\bullet$ {\em Anomalous diffusion}. The presented results could also apply whenever states of a statistical system are excluded by 
the presence of other materials restricting mobility in Euclidean space. Think e.g. of diffusion in porous media where statistically 
relevant action takes place on restricted surface-like areas, and not in full 3D.  

For non-commutative variables  alternative routes to generalized entropies may exist  \cite{Caruso2008,Saguia2010}. 

This work was inspired  by Prof. C. Tsallis to whom we are deeply indebted for countless discussions, suggestions and his 
fantastic hospitality at CBPF. We acknowledge  support by Faperj and CNPq (Brazilian agencies).


\end{document}